\documentclass[conference]{IEEEtran}
\IEEEoverridecommandlockouts
\usepackage{ifpdf}
\ifpdf

\usepackage{cite}

\ifCLASSINFOpdf
  \usepackage[pdftex]{graphicx}
  \DeclareGraphicsExtensions{.pdf,.jpeg,.png}
\else
  \usepackage[dvips]{graphicx}
  \DeclareGraphicsExtensions{.eps}
\fi
\usepackage{amsmath}
\usepackage{amssymb}
\usepackage{algorithm}

\usepackage[noend]{algorithmic}

\usepackage{array}
\ifCLASSOPTIONcompsoc
 \usepackage[caption=false,font=normalsize,labelfont=sf,textfont=sf]{subfig}
\else
 \usepackage[caption=false,font=footnotesize]{subfig}
\usepackage{url}
\usepackage{xurl}

\usepackage{xcolor}

\begin{document}
%
\title{FSD-Inference: \underline{F}ully \underline{S}erverless \underline{D}istributed Inference with Scalable Cloud Communication}

\author{\IEEEauthorblockN{Joe Oakley}
\IEEEauthorblockA{Department of Computer Science\\
University of Warwick\\
Email: J.Oakley@warwick.ac.uk}
\and
\IEEEauthorblockN{Hakan Ferhatosmanoglu*}
\IEEEauthorblockA{Department of Computer Science\\
University of Warwick\\
Email: Hakan.F@warwick.ac.uk\\
\thanks{*Also with Amazon Web Services. This publication describes work performed at the University of Warwick and is not associated with Amazon.}
}}


%


\maketitle

\begin{abstract}

Serverless computing offers attractive scalability, elasticity and cost-effectiveness. 
However, constraints on memory, CPU and function runtime have hindered its adoption for data-intensive applications and machine learning (ML) workloads. 
Traditional `server-ful' platforms enable distributed computation via fast networks and well-established inter-process communication (IPC) mechanisms such as MPI and shared memory. 
In the absence of such solutions in the serverless domain, parallel computation with significant IPC requirements is challenging. 
We present FSD-Inference, the first fully serverless and highly scalable system for distributed ML inference. 
We explore potential communication channels, in conjunction with Function-as-a-Service (FaaS) compute, to design a state-of-the-art solution for distributed ML within the context of serverless data-intensive computing.
We introduce novel \textit{fully serverless} communication schemes for ML inference workloads, leveraging both cloud-based publish-subscribe/queueing and object storage offerings. We demonstrate how publish-subscribe/queueing services can be adapted for FaaS IPC with comparable performance to object storage, while offering significantly reduced cost at high parallelism levels.
We conduct in-depth experiments on benchmark DNNs of various sizes. The results show that when compared to server-based alternatives, FSD-Inference is significantly more cost-effective and scalable, and can even achieve competitive performance against optimized HPC solutions. Experiments also confirm that our serverless solution can handle large distributed workloads and leverage high degrees of FaaS parallelism.


\end{abstract}
\section{Introduction}
\label{s:introduction}

Besides its popular stand-alone usage, ML inference is embedded in every step of data-intensive applications. There are a range of options for hosting large-scale inference workloads, including GPU clusters and HPC platforms. 
Such systems typically offer fast connectivity and IPC solutions such as MPI (Message Passing Interface)~\cite{Walker1995} and shared memory. 
These platforms are suitable for low latency and high utilization workloads, but are expensive to commission and maintain. 
\textbf{Server-based} \textit{cloud} compute resources are typically provisioned in two ways; `\textbf{job-scoped}' (i.e., starting resources for the duration of a processing job, and then shutting them down), or `\textbf{always-on}'. 
The startup time under the former (often several minutes) would rule out interactive inference query performance.
In the always-on case, while response times may meet requirements (although not necessarily for provider-managed inference endpoints~\cite{Wu2021}), costs can be high~\cite{Muller2020Lambada}. 

\textbf{Serverless computing} is a resource delivery model in which the cloud provider is responsible for the provision and management of underlying infrastructure and services. A core serverless offering is Function-as-a-Service (\textbf{FaaS}), which allows users to build and run applications as functions in stateless containers. Commercial FaaS offerings include AWS Lambda~\cite{AWSLambdaGeneral}, Google Cloud Functions~\cite{GoogleCloudFunctionsGeneral} and Azure Functions~\cite{AzureFunctionsGeneral}. The attractive properties of serverless computing include elasticity, cost-effectiveness with granular billing, and high availability. However, FaaS also presents challenges beyond those typically faced when using provisioned compute resources. These include capacity limits imposed by the cloud providers (e.g., on runtime and memory), a lack of direct inter-function communication, and `cold starts', leading to delayed invocation. FaaS is thus limited by its design and service constraints when applied to (distributed) data-intensive workloads.
Novel serverless communication and data management solutions are needed to address these challenges and adapt serverless for a wider range of data-intensive use cases.

Serverless computing has recently been identified as a viable option for ML inference~\cite{Gillis2021, Jarachanthan2021AMPS, Park2022AYCI}.
Cloud providers have introduced commercial \textbf{serverless inference} solutions. AWS offers SageMaker Serverless Inference~\cite{AWSServerlessInference} based on its Lambda FaaS service. Azure Databricks Serverless Real-Time Inference is a similar solution. While these products act as alternatives to provisioned endpoints for serving inference requests, they are subject to the inherent challenges summarized above. 
For example, each request is executed by a single, resource-constrained FaaS instance, limiting the model size. Large models and input data are increasingly utilized, due to their improved predictive performance. Hence, a distributed ML solution is required, as the models are bigger than can be accommodated in a single instance's memory. The significant compute, memory and inter-process communication (IPC) requirements associated with processing such models makes it challenging to apply FaaS to this popular use case.

In this paper, we present \textbf{FSD-Inference}, the first \textit{fully serverless} scalable solution for distributed ML inference that can also exploit sparsity in the underlying data and employ effective point-to-point communication. We achieve \textit{intra-layer model parallelism} to mitigate the limited memory of FaaS instances, utilize many concurrent function invocations to scale out compute, and introduce fully serverless IPC schemes. The latter enables the required data-intensive function-to-function interactions associated with inference over large and deep ML models. FSD-Inference leverages the benefits of cloud \textit{publish-subscribe/queueing and object storage} offerings with a smart communication mechanism. 
We also provide a \textit{rigorous cost model} for FSD-Inference, which informs our serverless inference design recommendations. 
Experiments confirm that our solution achieves low latency and high throughput, while scaling to large data workloads at low cost. We identify pub-sub/queueing as the most cost-effective approach as compute parallelism increases, object storage as the leading choice for very large inference tasks, and serial FaaS execution as the preferred option for small models/inference batch sizes.


In situations where inference workloads are \textbf{sporadic} in nature or varying in requirements (e.g., large/small models, few/many inference samples), neither an always-on, server-based platform, nor a commercial single-instance FaaS endpoint, are good fits. Müller et al.~\cite{Muller2020Lambada} describe data analytics settings where ad-hoc, interactive queries are performed sporadically. 
Analogous inference use cases are common, where queries are irregular, and require support for flexible model/input sizes. 
Further, sporadic requests can often be buffered before being processed as a \textbf{batch}. 
In such a workload, the trigger point for initiating inference may be unpredictable, and can occur in periods of bursty traffic or in response to a dynamic event. A serverless solution may be more suitable here than pre-provisioned resources, due to its short start-up time, scalability, and low cost. 
However, performing \textbf{batch inference} brings the additional challenges of scaling up the compute, memory and IPC capacities. Our solution therefore exploits data parallelism to efficiently process large batch inference requests concurrently, and supports larger batch sizes than commercial serverless offerings~\cite{FN-Serverless-Endpoints}. 



We further improve our solution's performance with \textbf{additional optimizations}. To mitigate I/O bottlenecks, we design a novel approach which distributes communication requests over multiple instances of cloud resources (e.g., publish-subscribe topics, object storage containers), and overlaps communication with computation. 
We develop an efficient point-to-point communication mechanism to transmit only the required intermediate results to each target processor in each layer. 
Given the limited compute resources of FaaS instances, we offload processing (e.g., message filtering and distribution) to back-end cloud services. 
Our solution aims both to minimize the time taken to construct and publish message payloads, and to maximize the utilization of provider-restricted capacities. 
Another optimization is our model partitioning strategy for distributed ML. We adapt hypergraph partitioning~\cite{Demirci2021} in this new context of FaaS-based distributed inference, and show its effectiveness when compared to simpler partitioning schemes. 
The goals of partitioning (i.e., reducing data volumes shared between processors and balancing their workloads) align well with the constraints inherent in model-parallel serverless ML. 
Finally, within each function instance, the use of compute resources is optimized via multi-threading to parallelize non-dependent processes such as message publication.

There are no prior serverless inference solutions with \textbf{fully serverless communication} (vs relying on serverful components) and \textbf{intra-layer model parallelism} (vs simpler layer-wise partitioning). Our FaaS-based engine provides a high-performance MPI-style platform for distributed ML inference, using off-the-shelf components. It features fine-grained data/model-parallelism suitable for many architectures, including NLP models/LLMs. 
Our main contributions are:
\begin{itemize}
    \item FSD-Inference is the first fully serverless and highly scalable system for distributed ML inference. 
    \item We design novel serverless point-to-point communication schemes, using both pub-sub/queueing and object storage services. These are applicable to many cloud-based machine learning and data-intensive applications. 
    \item We introduce a hierarchical function launch mechanism to minimize startup delays and enable instances to automatically determine their position in the execution tree.
    \item We formalize and validate a cost model for distributed serverless ML inference, and offer optimized design recommendations for workloads of various scales.
    \item Our solution can effectively exploit sparsity, achieves both model and data parallelism, and is further improved via hypergraph partitioning across FaaS instances. No prior work addresses \textit{intra-layer parallelism} of model architectures for scalable serverless machine learning.
    \item We investigate the communication patterns of cloud-based pub-sub/queueing and object storage IPC channels, and categorize the workloads where each solution would be most performant and cost-effective.
    \item We experimentally compare FSD-Inference against several cloud/server-based alternatives and an optimized HPC solution, using benchmark DNNs of various sizes. Experiments show that our FaaS-based solution achieves significant scalability, as well as an attractive cost-to-performance ratio.
\end{itemize}


The rest of this paper is organized as follows. Section~\ref{s:designing-cloud-ml-solution} introduces the key building blocks of a distributed serverless inference system. Section~\ref{s:fds-inf-design} presents the design of FSD-Inference, and its novel elements and methods. Section~\ref{s:cost-model} illustrates our cost model, and provides design recommendations for various ML and data-intensive workloads. Section~\ref{s:related-work} presents related work.  Section~\ref{s:experiments} describes the experimental evaluation. Finally, Section~\ref{s:conclusion} concludes the paper.
\section{Designing a Cloud-based Serverless ML Solution}
\label{s:designing-cloud-ml-solution}

In this section, we consider several key building blocks underpinning a scalable solution for serverless ML. These include the choice of compute engine, distributed processing and IPC patterns, model partitioning strategy, and communication channel design.

\begin{table*}[h!]
\centering
\caption{Features of potential inter-worker communication channels. Asterisks indicate partial support.}
\begin{tabular}{|l|c|c|c|c|c|c|c|}
    \hline
     & Stream & Stream (ETL) & NoSQL & Pub-Sub & Queues & \textbf{Pub-Sub+Queues} & \textbf{Object Storage} \\ \hline
     Serverless & \checkmark* & \checkmark & \checkmark* & \checkmark & \checkmark & \checkmark & \checkmark \\ \hline
     Low latency/high thrpt. & \checkmark & \checkmark & \checkmark & \checkmark & \checkmark & \checkmark & \checkmark \\ \hline
     Cost-effective & \checkmark* & \checkmark & & \checkmark & \checkmark & \checkmark & \checkmark* \\ \hline
     Flexible payloads/messages & & & & & & & \checkmark \\ \hline
     Many producers/consumers & \checkmark* & \checkmark & \checkmark & \checkmark & \checkmark & \checkmark & \checkmark \\ \hline
     Service-side filtering & & \checkmark & \checkmark & \checkmark & & \checkmark & \checkmark \\ \hline
     Direct consumer access & \checkmark & & \checkmark & & \checkmark & \checkmark & \checkmark \\ \hline
\end{tabular}
\label{tab:comm-channels}
\end{table*}

\subsection{Serverless Compute}
\label{ss:design-serverless-compute}

We target `scaled-by-request' FaaS offerings such as AWS Lambda and Azure Functions, where all aspects of infrastructure provision are taken care of by the cloud provider, and where each function instance runs in a dedicated, lightweight container. 
This has become a mature, widely-used technology with known performance characteristics, and has been the basis of other studies around serverless ML~\cite{Gillis2021, Jarachanthan2021AMPS}.

\subsection{Cloud-based Distributed ML}
\label{ss:design-cloud-ml}

There are many ways to enable parallel and distributed ML on cloud compute services. An ecosystem of open-source and commercial toolsets is available, aiming to simplify the running of parallel ML workloads~\cite{DistributedMLSurvey}. 
These solutions cater for a variety of hardware configurations; multi-CPU/multi-GPU architectures on single servers; multi-container configurations; multi-VM clusters and large HPC clusters. 
They range from generic interoperability/IPC solutions such as Message Passing Interface (MPI), to those more focused on data analysis and ML tasks, such as Ray~\cite{Ray2018}, Spark~\cite{Spark}, Kserve~\cite{KServe}, Dask~\cite{Dask}, and Horovod~\cite{Horovod}.
However, they all require direct node-to-node communication over (usually) TCP/IP.
In the case of `scaled-by-request' FaaS offerings, however, no such instance-to-instance communication is currently available, which means none of the above distributed computing services can be used.

Whilst interesting recent studies~\cite{Bohringer2022FMI, Wawrzoniak2021Boxer} work around the lack of direct FaaS inter-communication via mechanisms such as NAT hole punching, they come without provider support and hence lack SLOs and robustness guarantees. Our approach is instead to build a distributed communication solution using off-the-shelf services. Our objectives are:

\begin{enumerate}
    \item Construction of a flexible invocation tree of co-operating FaaS instances.
    \item For the required tree of FaaS instances to be launched as quickly as possible, by spreading the responsibility for starting instances across all internal nodes (each such node invokes its sub-tree of instances as a precursor to executing its compute role).
    \item For each instance to have a `rank' (in MPI terminology) and to understand its place in the tree (root/internal node/leaf).
    \item For each instance to understand which other instances it must communicate with in each layer. 
    \item For each instance to be able to communicate with its required targets via a high-throughput, fully serverless communication channel.
    \item For a nominated root instance to coordinate collective data transfer operations of various types (Barrier, Reduce, AllReduce etc.)
\end{enumerate}

\subsection{Model Partitioning in FaaS Settings}
\label{ss:design-partitioning}

Since serverless functions are constrained in terms of memory and compute, we need to partition the per-layer weights and inputs of large ML models across FaaS workers. This would also help balance the computational workload while minimizing the overall communication volume.
A variety of partitioning schemes have been successfully employed on traditional HPC systems.
However,
in contrast to HPC settings, serverless brings challenges of slower invocation/communication, resource-constrained instances and the requirement to reload large model weights for each request. Hence, it is particularly important to develop an effective partitioning scheme in this new context, in order to reduce both cost and run-time jointly.

\subsection{Serverless Communication}
\label{ss:design-serverless-communication}
There are currently no native and direct serverless FaaS-to-FaaS communication solutions analogous to the mechanisms used in the HPC setting. 
Reasons for this include function isolation and fast container switching \cite{Agache2020Firecracker, Brooker2023LambdaContainers}. 
To support distributed serverless inference, an ideal inter-function cloud communication channel should have the following properties:

\begin{itemize}
    \item Serverless (i.e., elasticity, pay-as-you-use, rapid automatic scale up/down)
    \item Low latency and high aggregate throughput
    \item Cost effectiveness
    \item Flexible payload type and message size
    \item Scalable to many concurrent producers and consumers
    \item Service-side message filtering
    \item Direct access to messages for consumers
\end{itemize}

We consider here a range of cloud service categories which could potentially be adapted to support the IPC element of a \textbf{fully serverless} distributed ML solution.

\subsubsection{Data Streaming Systems}

These solutions offer high throughput and low latency event logs.
For an ML inference workload, parallel FaaS workers could act as both producers and consumers. 
A stream is usually partitioned in such a way that it can be distributed across multiple processing nodes.
Commercial cloud data streaming offerings include AWS Kinesis Data Streams, Azure Event Hubs and Google Datastream.

\subsubsection{ETL on Streaming Data}
Cloud providers also offer extract-transform-load (ETL) solutions on streaming data (e.g., AWS Kinesis Data Firehose). Transformation features include data disaggregation and filtering, key-based partitioning, and compression.

\subsubsection{NoSQL Databases}
These services typically offer low latency and high scalability (often auto-scaling based on read and write throughput). Several design options could be developed for an ML inference workload, including the maintenance of a `global' intermediate output vector/matrix, as well as a message passing approach. 

\subsubsection{Publish-Subscribe}
Pub-Sub services offer a simple communication approach in which messages are distributed from any number of producers to unlimited subscribing consumers. 
Service-side filtering policies reduce the computational load on parallel consumers, which can be particularly beneficial in resource-constrained FaaS use cases. 

\subsubsection{Message Queues}
Cloud-based message queues offer low latency and cost-effective enqueue and dequeue functionality, as well as adjustable polling settings to cater for varying message arrival rates. They can also act as a destination for pub-sub services in a `fan-out' design. 

\subsubsection{Object Storage}
Object storage offers immutable and scalable repositories catering for very large individual file sizes. Such services provide high aggregate throughput, when used in conjunction with parallel FaaS instances~\cite{Jonas2017Occupy}. 

Table \ref{tab:comm-channels} illustrates that, of the offerings described above, the categories most suitable for our use case are \textbf{publish-subscribe (pub-sub) in conjunction with message queues}, and \textbf{object storage}. 
They are both fully serverless, with rapid scale up and down, and no \textit{`hangover costs'} (i.e., slow scaling down of burst-provisioned capacity units). 
They also offer flexibility of FaaS parallelism without requiring reconfiguration of communication resources, as well as \textit{service-side message filtering} to reduce the computational load on parallel workers. 

While any of the other services above \textit{could} potentially be adapted for a distributed ML solution, each has one or more significant shortcomings for our use case.
Data streaming offerings can be restrictive in terms of the numbers of producers/consumers allowed per provisioned resource, in time lag when auto-scaling these resources, and in the maximum number of API calls per second~\cite{FN-AWS-Stream-Quotas, FN-Microsoft-Event-Hubs-Quotas, FN-Google-Datastream-Quotas}.
Streaming ETL solutions do not offer \textit{direct polling} of the delivery stream, which adds an additional latency overhead as data must instead be written to/read from an intermediate location. They also have large minimum buffer sizes (both in terms of data volume and time triggers). 
NoSQL databases suffer from restricted item sizes, limited batch update functionality, and relatively high cost compared to other  offerings~\cite{FN-AWS-DynamoDB-Quotas, FN-Azure-CosmosDB-Quotas}.

In the context of serverless ML inference, we will examine the trade-off between the lower cost but payload-restrictive messaging supported by pub-sub/queueing based solutions, with the more expensive but effectively size-independent data transfers possible when using object storage.

\section{FSD-Inference Design}
\label{s:fds-inf-design}


FSD-Inference (\underline{F}ully \underline{S}erverless \underline{D}istributed Inference) is the first fully serverless system for distributed deep neural network (DNN) inference. 
One of the first challenges is the efficient 
parallelization of fully-connected layers across multiple FaaS instances, due to the large communication requirements and lack of direct FaaS-to-FaaS connectivity. We therefore introduce two fully serverless point-to-point data communication solutions for the cloud. 
The overall solution mitigates the constraints of FaaS and achieves scalable performance for ML inference at low cost, also maintaining its high performance under sporadic workloads.

\begin{figure}[h]
\centering  
\includegraphics[width=0.45\textwidth]{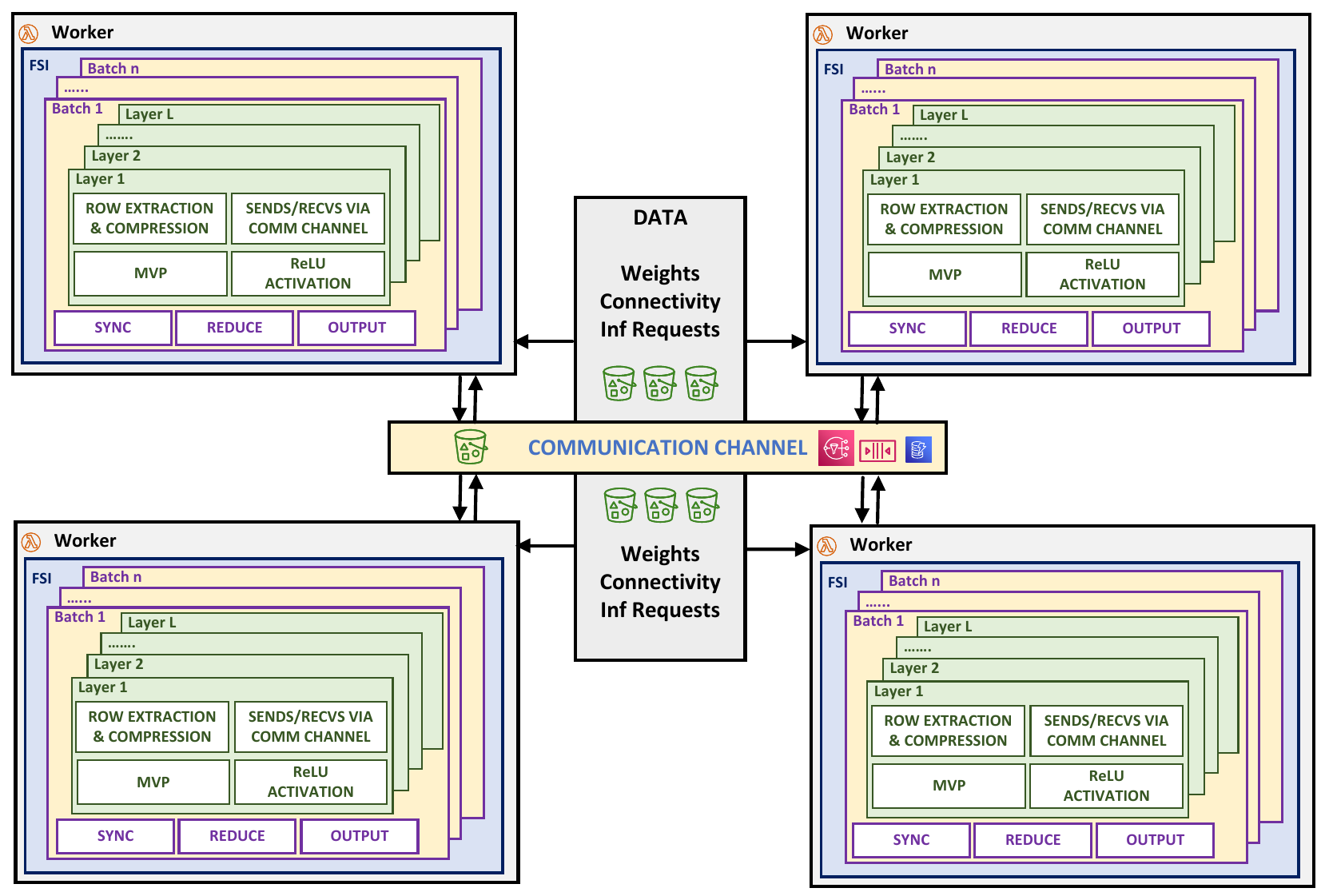}
\caption{FSD-Inference high-level cloud architecture}
\label{fig:overall-architecture}
\end{figure}


Our solution is cloud-provider agnostic. In light of AWS Lambda’s short `cold start' time, relatively long maximum runtime (15 minutes) and best-in-class performance~\cite{Rifai2021, ServerlessSurvey2022}, we present our design principles for FSD-Inference on AWS Lambda. Hence, we also adapt AWS services for IPC channels; namely, Simple Notification Service (SNS) and Simple Queue Service (SQS) for pub-sub/queueing, and Simple Storage Service (S3) for object storage. We characterize FSD-Inference with these two communication approaches as \textbf{FSD-Inf-Queue} and \textbf{FSD-Inf-Object}, respectively. It should be stressed that equivalent solutions could readily be built on other cloud platforms. Note that we also consider the use of FSD-Inference with a single worker with no communication components; we term this variant \textbf{FSD-Inf-Serial}.

The tree structure of parallel FaaS instances is built using a \texttt{worker\_invoke\_children()}method, in which a worker uses its parent-ID, `sibling number' and a branching factor to determine its own ID (similar to `rank' in MPI), and then to invoke its children. 
This process enables our point-to-point communication schemes, and allows workers to identify the weights/data partitions it should retrieve prior to commencing inference.
Experiments (not shown) also indicate that this mechanism reduces the launch time for the fully populated instance tree, compared to a centralized single-loop launch or a two-level launch loop as used in Lambada \cite{Muller2020Lambada}.

The solution is fully parameterized, meaning that at inference time, a request can be run with any number of workers $k$, provided that the model has been pre-partitioned offline for that $k$. The user can select different $k$ for consecutive requests, if desired. We consider offline hypergraph partitioning as post-processing of trained models (it is done a priori, not for each request).



Figure \ref{fig:overall-architecture} displays the high-level architecture of FSD-Inference. 
Upon invocation, each FaaS instance (worker) launches its children as above, then reads its share of the model weights, inference data and per-layer send and receive maps. 
Each worker then executes the FSI (Fully Serverless Inference) routines presented in Algorithms \ref{alg:sspinf-queue} (FSD-Inf-Queue) and \ref{alg:sspinf-object} (FSD-Inf-Object).
FSD-Inference introduces a highly parallel approach with an efficient communication mechanism to mitigate the inherent constraints of FaaS to achieve a scalable serverless solution.  

\subsection{FSD-Inf-Queue Communication Channel}
\label{ss:sds-inf-queue}



\begin{figure}[h]
\centering  %
\includegraphics[scale=0.4]{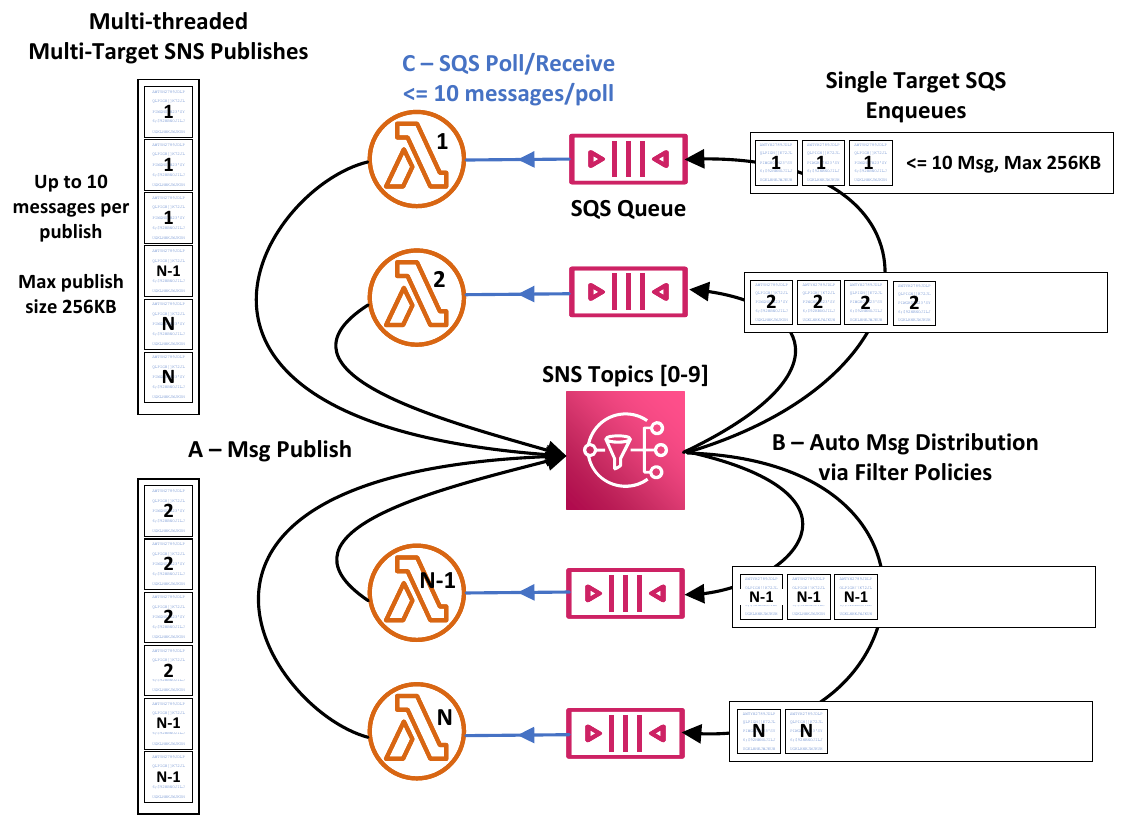}
\caption{FSD-Inf-Queue communication scheme. Demonstrated with AWS services (SNS, SQS, Lambda).}
\label{fig:sns-comm}
\end{figure}

Figure~\ref{fig:sns-comm} illustrates the FSD-Inf-Queue architecture. 
To design a fast and scalable serverless communication scheme for distributed ML inference, we aim to maximize the overall communication throughput, whilst avoiding I/O bottlenecks and minimizing the computational overhead on the resource-constrained FaaS instances. 
Several design features of FSD-Inf-Queue contribute to these goals. 
These include parallel pub-sub topics to increase total throughput and avoid I/O bottlenecks, and a dedicated message queue per FaaS instance to minimize the number of service-to-service network connections required. 
This also avoids the need for consumer-side message filtering (and hence the parsing/discarding of unwanted messages intended for other workers). 
We leverage a `fan-out’ architecture to offload the targeted distribution of messages onto the pub-sub service, helping to reduce computational overheads. 
Finally, the communication resources (pub-sub topics, queues, etc.) are pre-created to avoid doing so at inference time, at no additional provisioning or ongoing cost. 
Further details on how this communication scheme is utilized for ML inference are provided in Section~\ref{sss:pub-sub-queueing-inference-specifics}.

\subsection{FSD-Inf-Object Communication Channel}

\begin{figure}[h]
\centering  %
\includegraphics[scale=0.4]{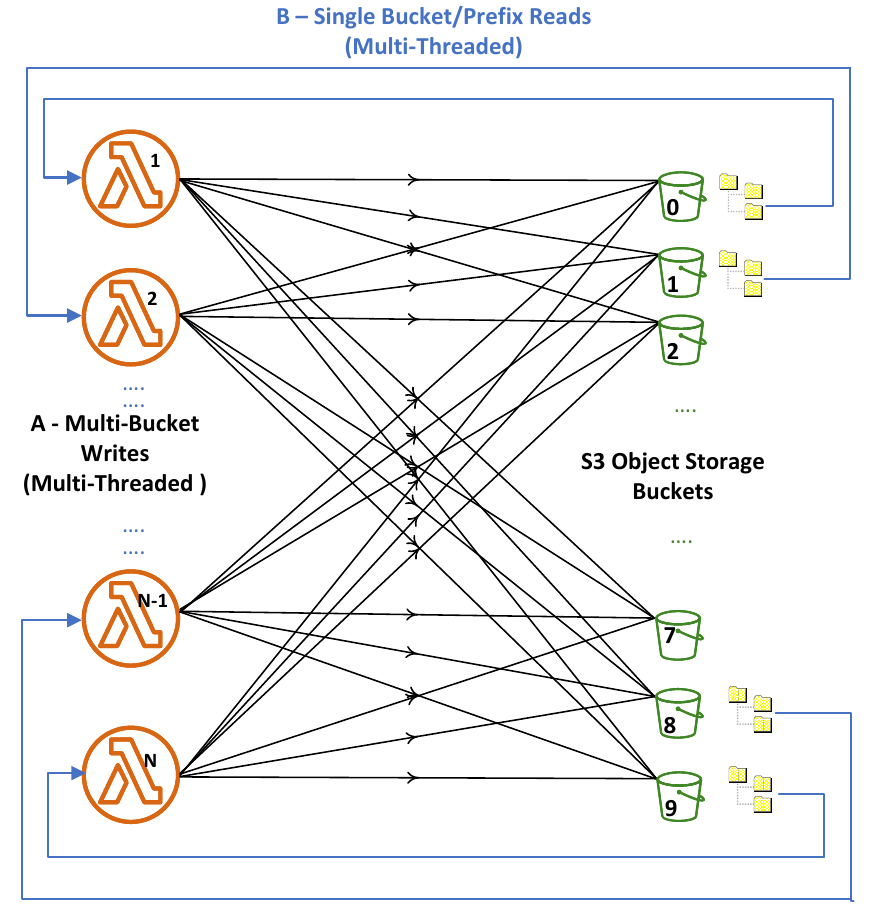}
\caption{FSD-Inf-Object communication scheme. Demonstrated with AWS services.}
\label{fig:s3-comm}
\end{figure}

As in FSD-Inf-Queue, our key objectives are to maximize communication throughput at minimal computational effort for FaaS workers. In FSD-Inf-Object, we employ multiple object containers and prefixes for the intermediate results files. This helps avoid I/O bottlenecks, whilst staying within provider-imposed API quotas~\cite{FN-AWS-S3-Performance-Optimization}; in object storage solutions such as S3, using $k$ containers effectively increases the API request limit $k$-fold \cite{Muller2020Lambada}. Under this design, each worker only needs to scan/read from a single object storage container and file prefix. The unnecessary reading of empty files (i.e., where a source had no information to transmit to a target in the current layer) is avoided by the use of distinct suffixes for empty/non-empty files. We also initiate object storage read operations in multiple parallel threads per worker, to minimize delays between folder scans. Figure~\ref{fig:s3-comm} presents the FSD-Inf-Object architecture, and Section~\ref{sss:object-storage-inference-specifics} provides further details on its use for ML inference.

\subsection{FSI Algorithm}
\label{ss:serverless-spff}


We propose two FSI (Fully Serverless Inference) algorithms, each performing repeated distributed MVP (matrix-vector product, for single sample inference) / MMP (matrix-matrix product, for batch inference) in the form of $W^kx^{k-1}$  / $W^kX^{k-1}$ for each layer $k=1,2,\dots,L$. Our inference algorithms and communication schemes are designed to exploit sparsity in DNN weight matrices/activations, but our ideas are readily applicable to dense inference scenarios. We will henceforth present notation for MVP, but this should be understood as also applying to MMP. Models are parallelized through the row-wise hypergraph partitioning of (sparse) weight matrices $W^k$ and input/output vectors $x^{k-1}$, which we adapt in this FaaS setting~\cite{Demirci2021}. 


Algorithms~\ref{alg:sspinf-queue} and \ref{alg:sspinf-object} describe the execution of the serverless FSI routine using the pub-sub/queueing (FSD-Inf-Queue) and object storage (FSD-Inf-Object) communication channels, respectively. In both algorithms, each FaaS processor $P_m$ (for $m = 1, 2, \dots, P$) holds row-blocks $W^k_m$ and $x^{k-1}_m$ of $W^k$ and $x^{k-1}$, respectively. The hypergraph partitioning scheme also provides each processor $P_m$ with maps $Xsend_m^k$ and $Xrecv_m^k$ which match row indices of $x^{k-1}_m$ to the target/source worker IDs with which they should be shared \cite{Demirci2021}. 


In order to perform MVP $W^k_mx^{k-1}_m$, worker $P_m$ must receive all rows of $x^{k-1}$ corresponding to the indices of the nonzero weight matrix columns it holds for layer $k$, from the workers which manage these rows.


The mechanism for sending and receiving intermediate results using FSD-Inf-Queue and FSD-Inf-Object is as follows:

\subsubsection{FSI with FSD-Inf-Queue}
\label{sss:pub-sub-queueing-inference-specifics}

For each target $P_n$ that source worker $P_m$ must communicate with in a given layer, we construct messages to send to a pub-sub topic. Due to the current provider-imposed constraints of the cloud pub-sub service, these messages consist of size-limited byte strings. We use the total number-of-nonzeros (NNZ), over the rows to be communicated, as a heuristic to estimate the number of byte strings required to encode $\bar{x}^{k-1}_{mn}$. This list of byte strings is denoted as $\{\bar{x}^{k-1}_{mni}\}$. Our goal is to maximize the utilization of the allowed pub-sub message size, whilst grouping and compressing rows only once.

All byte strings are added to a send buffer $Xsend\_list^k_m$, from which we populate a series of message \textit{batches} (to reduce API calls and hence cost); this process is multi-threaded in each worker. Note that we also add several message attributes, including the source worker ID, the total number of byte strings to be sent to this target, and the message layer. 
Having constructed these message batches, we then use parallel threads to issue them to the pub-sub topic, which distributes them to each target's dedicated message queue using a filter policy. We then perform local MVP $z^k_m = W^k_mx^{k-1}_m$ to overlap communication with computation (amortizing communication overheads), before receiving inbound $x^{k-1}$ rows from other workers.

 \renewcommand{\algorithmicrequire}{\textbf{Input:}}

\begin{algorithm}
\caption{FSI with FSD-Inf-Queue}
\begin{algorithmic}[1]
\label{alg:sspinf-queue}
\REQUIRE $x^0_m, \{W^k_m \}, \{ Xsend^k_m \}, \{ Xrecv^k_m$ \}
\FORALL{FaaS workers $P_m$ in parallel}
    \FOR{$k = 1, 2, \dots, L$}
        \FORALL{$(P_n, \bar{x}_{mn}^{k-1}) \in Xsend_m^k$}
            \STATE $\bar{x}^{k-1}_{mn} \gets $\texttt{ extract\_rows($x^{k-1}_m, \bar{x}^{k-1}_{mn}$)}
            \STATE $Xsend\_list_m^k \gets Xsend\_list_m^k \cup \{(P_m, \bar{x}_{mni}^{k-1}) \} $
        \ENDFOR
        \FORALL{b $\in$ \texttt{pop\_batches}$(Xsend\_list_m^k)$}
            \STATE \texttt{topic-\{m\%10\}.publish\_batch(b)}
        \ENDFOR
        \STATE $z^k_m \gets W_m^kx_m^{k-1}$
        \WHILE{$len(Xrecv_m^k) > 0$}
            \STATE $Xpoll_m^k \gets \texttt{queue-\{m\}.poll()}$
            \FORALL{$(P_n, \hat{x}_{nmi}^{k-1}) \in Xpoll_m^k$}
                \STATE $Xrecv\_list_{m}^k \gets Xrecv\_list_{m}^k \cup \hat{x}_{nmi}^{k-1}$
                \IF{$Xrecv\_list_{m}^k[P_n] == Xrecv_{m}^k[P_n]$}
                    \STATE $Xrecv_{m}^k \gets Xrecv_{m}^k / (P_n, \hat{x}_{nm}^{k-1})$ 
                \ENDIF
            \ENDFOR
            \STATE \texttt{queue-\{m\}.delete($Xpoll_m^k$)}
        \ENDWHILE
        \FORALL{$\hat{x}_{nm}^k \in Xrecv\_list_m^k$}
            \STATE $z_m^k \gets z_m^k + W_m^k \hat{x}_{nm}^k$
        \ENDFOR
        \STATE $x_m^k = f(x_m^k)$
    \ENDFOR
    \STATE \texttt{barrier($P_{all}$)}
    \STATE \texttt{reduce($P_0, x_m^L$)}
\ENDFOR
\IF{FaaS Worker $P_0$}
    \RETURN $x^L$
\ENDIF
\end{algorithmic}
\end{algorithm}





In order to receive messages from other workers (lines 9 to 15 in Algorithm~\ref{alg:sspinf-queue}), worker $P_m$ repeatedly polls its queue until all rows in $Xrecv_m^k$ have been obtained. 
We poll the message queue using `long' polling. The AWS message queueing solution caters for `long' and `short' polling, determined via a parameter $W$. `Short' polling ($W=0$) responds to polls immediately, regardless of whether or not messages were found. Note that SQS distributes messages over multiple servers; `short' polling does not guarantee that all servers are queried, and so messages may be missed by a request. `Long' polling ensures that all servers are visited. If messages are not immediately found, the service waits for up to $W$ seconds (continuously checking for messages at no extra cost), and can only return empty-handed after this period has elapsed. Our analysis (not shown in this paper) confirms that long polling outperforms short polling, and returns significantly more messages per poll request, reducing costs.

For each message returned in the poll, we add its deserialized and decompressed body to $Xrecv\_list_{m}^k$. We cater for the case where source $P_n$ needs to send multiple messages to target $P_m$ using message attributes, and only remove worker $P_n$ from $Xrecv_m^k$ when all its expected byte strings for layer $k$ have been received.


\subsubsection{FSI with FSD-Inf-Object}
\label{sss:object-storage-inference-specifics}


Due to the large file sizes available with object storage solutions, each FaaS instance in FSD-Inf-Object only needs to write a single object for each of its targets in a given layer (in contrast, FSD-Inf-Queue experiences far more constrained message sizes). 
Before performing non-blocking sends (lines 5/8 in Algorithm~\ref{alg:sspinf-object}), each worker iterates through tuples $(P_n, \bar{x}^{k-1}_{mn})$ in its send map for the current layer. For each row in $\bar{x}^{k-1}_{mn}$, we check if $P_m$ has any nonzero entries for that row. We maintain two sublists of row indices in $\bar{x}^{k-1}_{mn}$, indicating for which rows $P_m$ has nonzero entries and for which rows it does not.
Worker $P_m$ sending intermediate inference results to worker $P_n$ in layer $k$ would write a file named:

\texttt{s3://bucket-\{n\%10\}/\{k\}/\{n\}/\{m\_n\}.dat}

When $P_m$ has no nonzero entries to communicate to $P_n$ in layer $k$, it will instead write a 0-bytes file:

\texttt{s3://bucket-\{n\%10\}/\{k\}/\{n\}/\{m\_n\}.nul}

 \renewcommand{\algorithmicrequire}{\textbf{Input:}}

\begin{algorithm}
\caption{FSI with FSD-Inf-Object}
\label{alg:sspinf-object}
\begin{algorithmic}[1]
\REQUIRE $x^0_m, \{W^k_m \}, \{ Xsend^k_m \}, \{ Xrecv^k_m$ \}
\FORALL{FaaS workers $P_m$ in parallel}
    \FOR{$k = 1, 2, \dots, L$}
        \FORALL{$(P_n, \bar{x}_{mn}^{k-1}) \in Xsend_m^k$}
            \IF{$\bar{x}_{mn}^{k-1} == \emptyset$}
                \STATE \texttt{ObjStore.put\_obj("m\_n.nul", $\emptyset$)}
            \ELSE
                \STATE $\bar{x}^{k-1}_{mn} \gets $\texttt{ extract\_rows($x^{k-1}_m, \bar{x}^{k-1}_{mn}$)}
                \STATE \texttt{ObjStore.put\_obj("m\_n.dat", $\bar{x}^{k-1}_{mn}$)}
            \ENDIF
        \ENDFOR
        \STATE $z_m^k \gets W_m^kx_m^{k-1}$
        \WHILE{$len(Xrecv_m^k) > 0$}
            \STATE $Xhandles^k_m \gets $ \texttt{ObjStore.list\_files()}
            \FORALL{$h^k_{nm} \in Xhandles^k_m$}
                \IF{$h^k_{nm}$\texttt{.ends\_with(".nul")}}
                    \STATE $Xrecv^k_m \gets Xrecv^k_m / (P_n, \hat{x}^{k-1}_{nm})$
                    \STATE \texttt{continue}
                \ELSIF{$n \notin Xrecv^k_m$}
                    \STATE \texttt{continue}
                \ELSE
                    \STATE $\hat{x}^{k-1}_{nm} =$\texttt{ ObjStore.get\_obj($h^k_{nm}$)}
                    \STATE $Xrecv\_list^k_m \gets Xrecv\_list^k_m \cup \hat{x}^{k-1}_{nm}$
                    \STATE $Xrecv^k_m \gets Xrecv^k_m / (P_n, \hat{x}^{k-1}_{nm})$
                \ENDIF
            \ENDFOR
        \ENDWHILE
        \FORALL{$\hat{x}_{nm}^k \in Xrecv\_list_m^k$}
            \STATE $z_m^k \gets z_m^k + W_m^k \hat{x}_{nm}^k$
        \ENDFOR
        \STATE $x_m^k = f(x_m^k)$
    \ENDFOR
    \STATE \texttt{barrier($P_{all}$)}
    \STATE \texttt{reduce($P_0, x_m^L$)}
\ENDFOR
\IF{FaaS Worker $P_0$}
    \RETURN $x^L$
\ENDIF
\end{algorithmic}
\end{algorithm}


Workers do not attempt to retrieve files with the \texttt{".nul"} extension when they are identified in folder scans. 



In order to retrieve data sent by source FaaS instances, a worker will repeatedly scan the file handles in the location “\texttt{bucket-\{m\%10\}/k/m/}”. For all file names 
identified by the scan, we remove $(P_n, \hat{x}^{k-1}_{nm})$ from $Xrecv^k_m$ for files ending “\texttt{.nul}”. Any “\texttt{.dat}” files from sources $P_n$ where $(P_n, \hat{x}^{k-1}_{nm}) \notin Xrecv^k_m$ are ignored, as data from $P_n$ has already been received. The remaining “\texttt{.dat}” files are then read into memory, decompressed and added to $Xrecv\_list^k_m$. $(P_n, \hat{x}^{k-1}_{nm})$ is then removed from $Xrecv^k_m$.


\subsubsection{Post-Communication Processing}
This processing is common to both communication channels.
Once data from all the row indices in $Xrecv_m^k$ is stored in $Xrecv\_list_m^k$, we iterate through each (sparse) matrix $\hat{x}^{k-1}_{nm} \in Xrecv\_list_m^k$ and perform operation $z^k_m \gets z^k_m + W^k_m\hat{x}^{k-1}_{nm}$, to complete the distributed MVP for layer $k$. The final operation in each layer is the application of an activation function $f(\cdot)$.

After layer $L$, we perform Barrier and Reduce operations to synchronize all workers, and communicate all output vectors $x^L_m$ to worker $P_0$. $P_0$ then aggregates them to calculate the overall inference result $x^L$. This entire process can then be repeated across multiple successive inference samples/batches.

\section{FSD-Inference Cost Model}
\label{s:cost-model}
In this section, we first formalize the cost models of FSD-Inf-Serial, FSD-Inf-Queue and FSD-Inf-Object. We then introduce several optimizations, observing their effect on the cost models. Finally, we give recommendations for the design of serverless ML inference solutions. It should be noted that while we henceforth refer to AWS terminology and pricing models, 
the design principles are cloud-provider agnostic.

A detailed cost model is important due to the cost-to-performance tradeoffs of serverless, and to show the viability of FSD-Inference as an alternative to provisioned solutions. 
While previous work~\cite{Muller2020Lambada} has evaluated the costs of object storage-based exchange operators, our cost model is the first to consider end-to-end distributed serverless inference expenses.

\subsection{Serverless Inference Cost Analysis}
\label{ss:formalizing-cost-model}
Overviews of the cost models of FSD-Inf-Queue, FSD-Inf-Object and FSD-Inf-Serial are shown in Equations \ref{eq-c-queue-high-level}, \ref{eq-c-object-high-level} and \ref{eq-c-serial-high-level}, respectively. The following sections explore each in detail.
\begin{gather}
    C_{Queue} = C_{\lambda} + C_{SNS} + C_{SQS} \label{eq-c-queue-high-level}\\
    C_{Object} = C_{\lambda} + C_{S3} \label{eq-c-object-high-level}\\
    C_{Serial} = C_{\lambda} \label{eq-c-serial-high-level}
\end{gather}

\subsubsection{FSD-Inf-Queue Cost Model Breakdown}
\label{sss:queue-cost-model}
We first consider $C_{\lambda}$, which is common to all variants. The costs of the pub-sub and queueing communication components will then be subsequently described.
$C_{\lambda}$ is given as:
\begin{equation}
    C_{\lambda} = PC_{\lambda(Inv)} + P\bar{T}MC_{\lambda(Run)}
    \label{eq-c-lambda}
\end{equation}

Where $P$ is the total number of Lambda instances, $\bar{T} = \frac{1}{P}\sum_{i=1}^{P}T_i$ is the average worker runtime (in seconds), and $128 \le M \le 10240$ (at present) is the memory allocated to each $P_m$.  $C_{\lambda(Inv)}$ is the static cost per Lambda invocation, and $C_{\lambda(Run)}$ is the cost per MB-second of Lambda runtime~\cite{FN-AWS-Lambda-Pricing}. 
Since Lambda vCPU allocation is proportional to the allocated memory, there is an inherent cost-to-performance trade-off when sizing Lambda functions~\cite{FN-AWS-Lambda-Performance-Optimization}.  
As FSD-Inf-Serial does not feature a communication channel, $C_{\lambda}$ makes up the entirety of its cost profile.

Recall that the workflow of FSD-Inf-Queue is as follows: Source workers publish messages to SNS topics (via SNS publish requests), which distribute them to SQS queues. Target workers then perform SQS API requests to poll their queue to receive inbound messages. With this in mind, $C_{SNS}$ and $C_{SQS}$ can be calculated as follows:
\begin{gather}
    C_{SNS} = SC_{SNS(Pub)} + ZC_{SNS(Byte)}\\
    C_{SQS} = QC_{SQS(API)}
\end{gather}

Where $S$ is the number of billed publish requests, $Z$ is the total number of bytes transferred between the pub-sub and queueing services, and $Q$ is the number of API calls to the queueing service. $C_{SNS(Pub)}$ is the cost per billed publish request, $C_{SNS(Byte)}$ is the cost per byte transferred from SNS to SQS, and $C_{SQS(API)}$ is the cost per SQS API request. 
Note that while the current maximum payload size for SQS is 256KB, publishes are billed in 64KB increments. Hence, a publish containing 256KB of data (spread across up to 10 messages) will be billed as 4 requests. In addition, AWS currently only charges for data transfer from SNS to SQS, not from Lambda to SNS or from SQS to Lambda, provided all resources reside in the same AWS region.

\subsubsection{FSD-Inf-Object Cost Model Breakdown}
\label{sss:object-cost-model}
The compute costs of FSD-Inf-Object ($C_{\lambda}$) are as described above. Its communication costs, $C_{S3}$, are calculated as:
\begin{equation}
    C_{S3} = VC_{S3(Put)} + RC_{S3(Get)} + LC_{S3(List)}
    \label{eq-c-s3}
\end{equation}

 In Equation \ref{eq-c-s3}, $V$ represents the number of object storage \texttt{PUT} requests, $R$ is the number of \texttt{GET} requests, and $L$ is the number of \texttt{LIST} requests. \texttt{PUT} and \texttt{GET} requests are billed irrespective of the size of the object being written/read.
\subsection{FSD-Inference Optimizations}
\label{ss:technical-optimizations}
Our optimizations in FSD-Inference aim to reduce the costs described in the above models. FSD-Inf-Queue benefits from our approach to maximizing the publish payload utilization when packing byte string lists $\{\bar{x}^{k-1}_{mni}\}$, hence minimizing $S$. Further, the long polling mechanism illustrated in Section~\ref{sss:pub-sub-queueing-inference-specifics} reduces $Q$ when compared to alternative polling approaches. The optimizations in Section~\ref{sss:object-storage-inference-specifics} improve the cost profile of FSD-Inf-Object (avoiding reading “\texttt{.nul}” files, or performing redundant reads of any “\texttt{.dat}” files where $(P_n, \hat{x}^{k-1}_{nm}) \notin Xrecv^k_m$); these changes both reduce $R$. Further, overlapping S3 reads with the write phases of other workers, as illustrated in Algorithm \ref{alg:sspinf-object}, reduces the number of \texttt{LIST} requests $L$.

Both FSD-Inf-Queue and FSD-Inf-Object utilize \texttt{ZLIB} compression 
to reduce the communication volume. While this directly reduces FSD-Inf-Queue communication costs (by reducing $Z, S$ and $Q$), the improved performance experienced under the reduced IPC load benefits both variants by reducing $\bar{T}$, and hence $C_{\lambda}$. Similarly, the overall communication volume is reduced via hypergraph partitioning of the model across parallel workers, improving several components of both cost models (particularly by reducing $S$ and $Z$ for FSD-Inf-Queue).

\subsection{Serverless Inference Design Recommendations}
\label{ss:design-recommendations}
We conclude this section with several design recommendations for fully serverless ML inference systems. For small models which can comfortably fit into the memory of a single FaaS instance, single-instance execution is the recommended approach (i.e., FSD-Inf-Serial) as it avoids any communication latency of FaaS IPC.
For models which will not fit into the memory of (or be efficiently processed by) a single FaaS instance, our distributed solution using either of the proposed serverless IPC channels is needed. 

FSD-Inf-Queue is a highly cost-effective solution as parallelism grows. This is especially true until multiple publishes are consistently required for each target, based on the cloud provider's maximum publish payload capacity. 
At present, API costs for $C_{SNS(Pub)}$ and $C_{SQS(API)}$ are $\approx 1$ OOM less than $C_{S3(Put)}$ and $C_{S3(List)}$. 
Further, a single SNS publish can currently contain up to 10 messages, potentially satisfying layer-wise communication requirements for 10 targets simultaneously. S3, on the other hand, always requires one \texttt{PUT} per target, if redundant read/processing times for shared files are to be avoided. 
Further, a single SQS poll can retrieve payloads from up to 10 source workers (vs 1 read per source with S3).
Hence, in best-case conditions, pub-sub/queueing API costs can be up to 2 OOM lower than corresponding object storage requests, subject to message counts and sizes.
Hence, \textit{pub-sub/queueing costs will grow much more slowly with increasing \textbf{worker parallelism} than object storage costs, for a given data volume}.

Object storage becomes the optimal communication approach when the data volumes begin to saturate pub-sub/queueing resources. 
Inter-worker messages communicated via object storage can be of almost unlimited size (TB scale), and current pricing models do not charge for data transfer between S3 and Lambda instances. 
Hence, \textit{object storage communication channels have a favorable cost profile as \textbf{data volumes} grow}.

\section{Related Work}
\label{s:related-work}

\subsection{Serverless Computing}
\label{ss:related-work-serverless-overview}

In recent years there has been increasing exploitation of cloud-based serverless solutions for different types of computing workloads~\cite{Shafiei2022, Schleier-Smith2021, Khandelwal2020Taureau, ServerlessSurvey2022, Yu2020Benchmark1, Kim2019Benchmark2}. 
The range of use cases now includes data management and analytics~\cite{Muller2020Lambada, Perron2020}, DBMS/DBaaS~\cite{Li2022DBaaS}, linear algebra~\cite{Shankar2020NumPyWrenSLA}, ML inference~\cite{Ishakian2018, Jarachanthan2021AMPS}, ML training~\cite{Jiang2021Demystifying} and end-to-end ML pipelines~\cite{Paraskevoulakou2021}. 
Recent work addresses some of the challenges arising from the use of serverless architectures for data-intensive applications.


\subsubsection{Inter-Function Communication}
The lack of direct instance-to-instance communication in the serverless environment is a key challenge. A common approach~\cite{Muller2020Lambada, Perron2020} has been to use object storage as an intermediate layer for sharing state/data and message passing, via the design of serverless exchange operators. As well as our optimized object storage-based channel, we also design and evaluate a \textbf{pub-sub/queueing} approach, and compare the two in-depth, in terms of cost, performance and scalability. 
Several works~\cite{Jiang2021Demystifying, kodandarama2019serfer} make use of distributed memory caching solutions (e.g., Memcached, Redis), which can provide lower latencies and higher throughputs than object storage, but unlike our solution, they are \textbf{not fully serverless}. 
A NAT `hole-punching' approach is proposed~\cite{Wawrzoniak2021Boxer, Moyer2021} which enables direct communication between FaaS function instances via an external provisioned proxy server. This is incorporated in a wider serverless communication solution which emulates a subset of MPI functionality~\cite{Bohringer2022FMI}. Our work also implements MPI primitives (Send, Recv, Broadcast, Reduce), but \textbf{avoids the use of an external provisioned server}, achieving fully serverless communication. One serverless distributed inference solution~\cite{Gillis2021} achieves coordinator-worker communication using a fork-join launching mechanism, passing data via invocation requests/responses, although limited payload capacities can hinder scalability, and inter-worker communication is not considered. Instead, our work focuses on the design of highly parallel communication channels for \textbf{inter-worker communication}, for the purpose of intra-layer parallelism.

\subsubsection{Container Control}
Established FaaS offerings such as AWS Lambda provide limited customizability of runtime containers, hindering some use cases. 
Over recent years an ecosystem around Docker-based, Kubernetes-managed cloud compute services has also developed, with Knative and similar solutions enabling FaaS offerings to be built on top of this stack (e.g., Google Cloud Run~\cite{GoogleCloudRun}, IBM Cloud Code Engine ~\cite{IBMCloudCodeEngine}).
Such FaaS services are usually `scaled-by-concurrency', enabling multiple function instances to share containers.
These solutions provide opportunities for container customization, such as hosting web servers and shared storage, which could be leveraged by the hosted functions for IPC.
Another new category of offerings incorporates `stateful functions'~\cite{Statefun, Cloudburst2020, Orleans2011, AzureDurableFunctions2021} based on elements of the Actor model~\cite{ActorModel1973}. These offer state management and IPC mechanisms. However, they typically rely on non-serverless infrastructure. While this is an area of fluid change, our approach can readily adapt to such emerging FaaS architectures.

\subsubsection{Cold Starts and Inconsistent Performance}
Methods are proposed~\cite{Daw2020Xanadu, Carver2019DAG, Carver2020WukongNew, Icebreaker2022} for quickly launching many function instances while reducing the `cold-start' overhead this can entail, for example by pre-analyzing task chains and pre-launching resource instances.
Our work uses a \textit{multi-level, hierarchical launch design} which can mitigate the cold start issue.
The issue of `straggler' worker instances is also addressed by various methods including pre-emptive read/write retries~\cite{Perron2020} and by the use of local error-correcting codes~\cite{Gupta2020StragglerLocalErrorCorrecting, Bartan2019StragglerPolar}.

\subsubsection{Hybrid Solutions}
Several studies opt to mitigate the challenges of serverless by unifying serverless and serverful components~\cite{GarciaLopezServerMix2019, Suresh2021ServerMore, AsyFunc2023}. Gupta et al.~\cite{ReliableTransactionsICDE2023} address the integration of edge devices with multi-cloud FaaS instances and secure data stores in a fault-tolerant manner. In contrast, our work addresses a \textbf{fully serverless} design objective.

\color{black}

\subsection{Serverless Data Analytics and ML}
\label{ss:related-work-serverless-da-ml}

\subsubsection{Serverless Data Analytics}

Lambada~\cite{Muller2020Lambada} presents efficient scan, sort and transfer operations for data analytics, using object storage as a communication channel. 
Whilst Lambada uses queues for sending short result messages, we show that a pub-sub/queueing approach can also be viable for large scale data shuffling operations (i.e., moving data across partitions held on distributed nodes).
We also leverage hypergraph partitioning to minimize communication requirements and ensure balanced workloads across parallel workers. 
Our pub-sub/queueing approach also offloads the distributions and filtering of messages to the back-end cloud service.

Other works explore the use of serverless for Linear Algebra (LA) workloads. 
PyWren~\cite{Jonas2017Occupy} provides a MapReduce-style capability for data analytics workloads. 
NumPyWren~\cite{Shankar2020NumPyWrenSLA} leverages PyWren and LAmbdaPACK (a domain-specific language for coordinating LA algorithm tasks). 
Unlike these works, which study dense linear algebra, we focus on scalable ML inference. We are also able to leverage sparse data structures and point-to-point communication techniques which exploit the sparsity of the underlying data, i.e. both weight matrices and input vectors. This improves computational performance while reducing IPC overheads. Further, NumPyWren~\cite{Shankar2020NumPyWrenSLA} only mimics a serverless runtime, whereas FSD-Inference 
uses a true, off-the-shelf, fully serverless architecture.

\subsubsection{Serverless Machine Learning Systems}
A number of recent works consider the suitability of serverless platforms for ML workloads. 
There has been a particular interest in serverless ML \textit{inference} use cases~\cite{Wu2021}. Park et al.~\cite{Park2022AYCI} consider optimal FaaS configurations for inference, but do not cater for per-request parallelism. 
Tetris and Photons~\cite{Tetris2022, Photons2020} adapt Kubernetes-based FaaS offerings to enable runtime and tensor sharing to reduce memory usage, whilst our work achieves this via model and data partitioning.
AWS now offers a commercial serverless inference endpoint as part of its SageMaker suite~\cite{AWSServerlessInference}. Whilst this service can scale upwards (currently to 6GB memory) and outwards (via auto-scaling in response to demand levels), it does not yet address the challenges of per-request data parallelism, or ML model parallelism, which is a requirement for very large DNNs. 
A small number of studies \textit{have} attempted \textit{distributed} serverless inference.
AMPS-Inf~\cite{Jarachanthan2021AMPS} partitions trained models in a layer-wise fashion over multiple FaaS instances, and then runs them sequentially. Gillis~\cite{Gillis2021} adopts `layer grouping' algorithms which co-locate layers which should be processed together, thus avoiding substantial inter-instance communication. It then performs `coarse-grained partitioning' by assigning segments of some layers to small numbers of sub-worker instances, launched synchronously in a fork-join manner; communication only occurs via invocation requests/responses.
In contrast, our work is the first to employ \textbf{intra-layer} (i.e., tensor) model parallelism in the serverless setting, distributing consecutive fully-connected layers over multiple disconnected FaaS instances.

Another important area of interest is \textit{batch} serverless inference, as there are real-world use cases where an immediate response to an inference request is not required, and where request arrival is sporadic and/or bursty. This type of workload is a sweet-spot for serverless inference, as it can often be more cost-effective than sample-by-sample processing. BATCH identifies an SLO-optimal FaaS deployment (batch size, memory config) to carry out batch inference~\cite{Ali2020Batch}. Ali et al.~\cite{Ali2022OptInfServ} propose a performance and cost estimator 
to efficiently batch together heterogeneous serverless inference requests. 
Hybrid server-based/serverless batch inference solutions, which use serverless to supplement IaaS resources when handling bursty workloads, have also been proposed~\cite{Zhang2019}.
In contrast, our solution caters for batch inference processing as part of the core serverless design, although we assume that requests have been buffered and batched by an appropriate mechanism prior to FSD-Inference processing.

\section{Experimental Analysis}
\label{s:experiments}

\subsection{Experimental Setup and Datasets}
We first compare FSD-Inference against two server-based baselines, an HPC solution, as well as a commercial serverless inference offering, before contrasting our two fully serverless communication channels. 
We consider three variants of FSD-Inference; \textbf{FSD-Inf-Serial} (which runs on a single FaaS instance without communication), \textbf{FSD-Inf-Queue} (publish-subscribe/queueing) and \textbf{FSD-Inf-Object} (object storage).

We run the experiments on the popular benchmark provided by the MIT/IEEE/Amazon Sparse Deep Neural Network Graph Challenge \cite{Kepner2019}.
The objective of this benchmark is to facilitate the controlled evaluation of distributed solutions for large and deep networks, which is our primary area of interest. 
Our approaches can readily be generalized to other models and/or prediction tasks.

We use $L=120$ layers for all experiments, and evaluate performance with DNNs of per-layer neuron counts $N$ = 1024, 4096, 16384 and 65536.  
For inference data, \textbf{we process batches of 10,000} 
\textbf{samples}, which are scaled to $32 \times 32$, $64 \times 64$, $128 \times 128$ and $256 \times 256$ in accordance with $N$. 
These samples are thresholded and flattened into column vectors to conform with the input layers of the synthetic (sparse) DNNs. We then prepare the data for batch inference by concatenating these column vectors into a matrix for consecutive distributed MMP. 
We confirm our inference results match the ground truths provided by the benchmark to ensure correctness. 
For all experiments, we report the median results of 3 runs.


\subsubsection{FaaS Worker Configuration}
All experiments are performed using AWS Lambda as the compute service. 
For FSD-Inf-Serial, we create a Lambda application which runs Algorithm~\ref{alg:sspinf-queue} with all communication steps removed, and loads the unpartitioned DNN model and inference data into memory. We allocate the current maximum allowed memory $M$ = 10240MB, to enable whole models to fit into memory until the neuron count makes this infeasible.
For FSD-Inf-Queue and FSD-Inf-Object, we create two Lambda applications per variant, a `coordinator' and a `worker'. The lightweight coordinator ($M$ = 128MB) parses user input and invokes the first layer of workers. 
The worker function is configured with the maximum allowed runtime limit (15 minutes at the time of writing),  and executes Algorithm~\ref{alg:sspinf-queue} (FSD-Inf-Queue) or~\ref{alg:sspinf-object} (FSD-Inf-Object). \texttt{Boto3} (the AWS SDK for Python) is used to interact with AWS communication services. 
We invoke concurrent workers $P$ = 8, 20, 42 and 62. We allocate $M$ = 1000, 1500, 2000 and 4000MB for $N$ = 1024, 4096, 16384 and 65536, respectively. We sized Lambda functions such that the partitioned model weights could fit into memory, and allowed a small overhead beyond this. We employ Python's \texttt{concurrent.futures.ThreadPoolExecutor()} in each worker to parallelize communication as discussed in Section \ref{ss:serverless-spff}. We built FSD-Inference using Python 3.8. ReLU was selected as the non-linear activation function $f(\cdot)$, biases of -0.30, -0.35, -0.40 and -0.45 were applied for $N$ = 1024, 4096, 16384 and 65536, respectively, and neuron activations were thresholded to 32 (as per the Graph Challange). The hypergraph partitioning is implemented offline using PaToH \cite{PaToH}.

\subsubsection{Server Configurations}

Our rationale for sizing our server-based baselines was to use the smallest server with \textbf{greater} total vCPU and memory than the sum of the equivalent FSD-Inference resources, to ensure fairness. Note that official figures on Lambda vCPU allocation per unit of memory are \textit{not} published by AWS, so we rely on external analysis~\cite{Lambda-Memory-vCPU}.
For Server-Always-On, we use AWS EC2 c5.12xlarge compute-optimized instances (48 vCPU, 96 GiB memory). 
For Server-Job-Scoped, we utilize suitably-sized instances for each neuron count $N$. Namely, for $N$ = 1024 and $N$ = 4096 c5.2xlarge (8 vCPU, 16 GiB memory), for $N$ = 16384 c5.9xlarge (36 vCPU, 72 GiB memory), and for $N$ = 65536 c5.12xlarge (48 vCPU, 96 GiB memory).
We use the same codebase as FSD-Inf-Serial for these experiments.
With these configurations, the relative disadvantages that FSD-Inference experiences vs job-scoped baselines in terms of memory and vCPU are as follows (for each neuron count). Memory: 0.625x, 0.625x, 0.556x, 0.833x. vCPU: 0.721x, 0.721x, 0.639x, 0.963x. 

\subsection{Baselines}
First, we implement two server-based approaches for cloud-based ML inference: Server-Always-On (large VMs capable of handling peak load, left running between queries) and Server-Job-Scoped (VMs of a suitable size for a given request, invoked upon request arrival and closed down after termination). 

We also consider H-SpFF \cite{Demirci2021} as a non-cloud based approach, the results of which were achieved on an on-premise distributed HPC platform. H-SpFF results reflect the processing of 60,000 samples, as opposed to 10,000 in all other experiments (such results were not available). Note that cost information is also not available for H-SpFF. 
Finally, we evaluate AWS SageMaker Serverless Inference \cite{AWSServerlessInference} (denoted \textbf{Sage-SL-Inf}), using a PyTorch implementation of FSD-Inf-Serial. This ran on Lambda endpoints with the maximum allowed memory (6GB) to make it as competitive as possible. However, we omit this approach from Section \ref{ss:expts-handling-sporadic-inference-workloads}; it could not load all models into memory, while its restrictive maximum payload size (6MB) and function runtime (60s)  prevented it from processing the required 10,000 samples for any model size. 
We also applied Gillis~\cite{Gillis2021} to partition the models, and found that it only created Lambda resources for a single function which handles all layers, i.e., no layer grouping or parallelization. This configuration is equivalent to FSD-Inf-Serial execution, and hence we discount it as an explicit baseline.

\begin{figure}[h]
\centering  %
\includegraphics[scale=0.3]{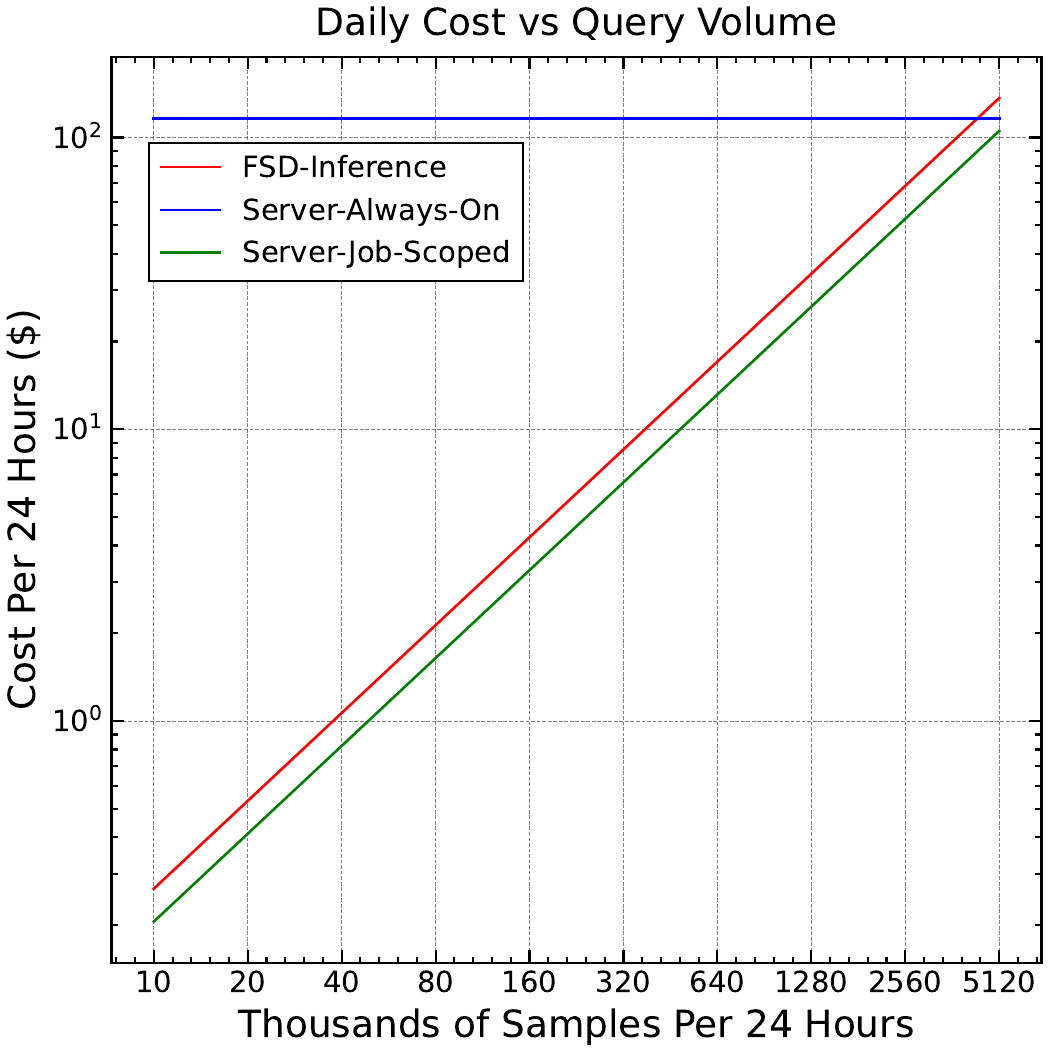}
\caption{Daily cost of FSD-Inference, Server-Always-On and Server-Job-Scoped for various query volumes. Queries are evenly spread between $N=1024, 4096, 16384$ and $65536$.}
\label{fig:cost-graph}
\end{figure}

\subsection{Handling Sporadic Inference Workloads}
\label{ss:expts-handling-sporadic-inference-workloads}



\subsubsection{Overview of Sporadic Inference Workloads}

We first consider the cost and performance of FSD-Inference, server-based baselines and H-SpFF on sporadic inference workloads. Under such a query pattern, requests may arrive at irregular and/or infrequent intervals, with varying model sizes. The execution of ML inference in response to a metric exceeding a threshold (e.g., in an e-commerce context, user retention time, no. of clicks on site, \% of basket items that lead to purchases) is a common example of such a use case. Other practical examples include financial trading, regional energy usage monitoring, real-time traffic forecasting, or social network analysis.
In these scenarios, the frequency and distribution of inference requests is unknown, and numerous different ML models could be invoked across queries.

\subsubsection{Modelling Sporadic Inference Workloads}
We model queries arriving over a 24-hour period, evenly spread over multiple per-layer neuron counts $N$. For each query, we select the FSD-Inference variant offering the best balance of cost and performance. For the always-on server baselines, we provision two of the specified instances, to handle overlapping queries and to offer redundancy. 
We also present `hot' and `cold' always-on results.
The former assumes that for 50\% of requests, the given ML model will already be in memory. In the other half of cases, the model is loaded from a high-performance EBS~\cite{FN-AWS-EBS} 
block storage volume, attached to the instance. In the case of `cold' results, the model is instead fetched from object storage.
This behaviour mimics SageMaker Multi-Model Endpoints~\cite{FN-AWS-Sagemaker-Multimodel-Endpoints}, 
which retain in memory the most recently-invoked models, with less in-demand models moved to EBS, and then S3, as capacity dictates.

\subsubsection{Cost and Performance Analysis}



\begin{figure}[h]
\centering  %

\includegraphics[width=0.43\textwidth]{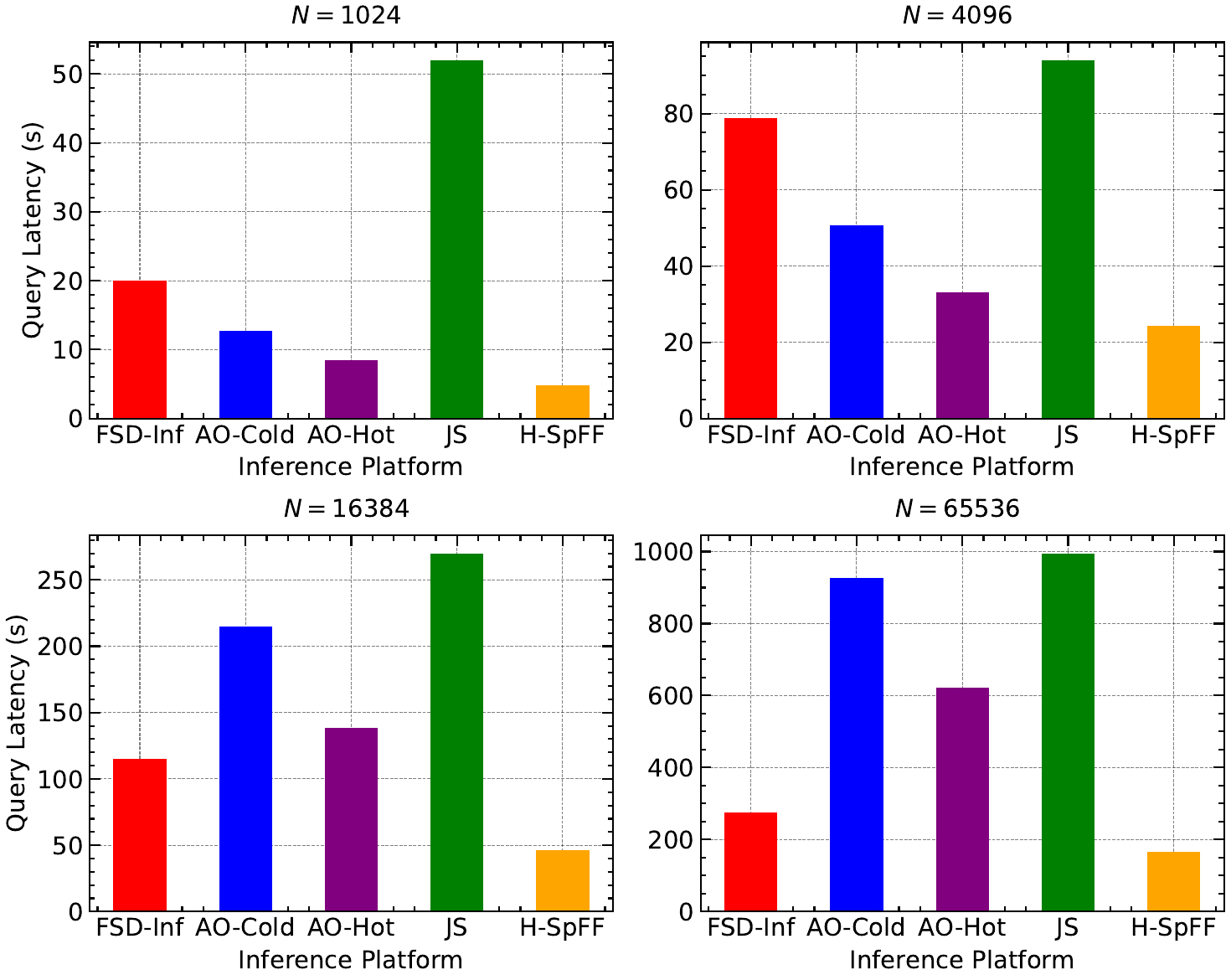}

\caption{Query latency for FSD-Inference, server-based baselines and H-SpFF.}
\label{fig:performance-graph-hot5050}
\end{figure}

As shown in Figure \ref{fig:cost-graph}, \textbf{FSD-Inference is significantly cheaper than Server-Always-On} until high daily query volumes are reached \textbf{(approx. 4M samples per day)}. While Server-Job-Scoped is marginally more cost-effective than FSD-Inference, it is seen in Figure \ref{fig:performance-graph-hot5050} that it suffers from \textbf{very high query latency for all model sizes}. For smaller models ($N = 1024, 4096$), FSD-Inference offers competitive performance compared to AO-Cold (Server-Always-On-Cold), but lags behind AO-Hot (Server-Always-On-Hot) and H-SpFF. This is in part due to overheads in reading unpartitioned weights from S3. For $N = 16384$, FSD-Inference is able to achieve superior performance compared to AO-Hot. Finally, for $N = 65536$, the scalability of FSD-Inference means it significantly outperforms AO-Hot, AO-Cold and JS, and has only $\approx 40\%$ higher latency than H-SpFF, an optimized HPC solution. These results show that serverless ML inference solutions can offer impressive scalability as well as an attractive cost-to-performance ratio, particularly for sporadic inference workloads.

\subsection{Analysis of Serverless Inference Platforms}

\begin{figure}[]
\centering  %
\includegraphics[scale=0.36]{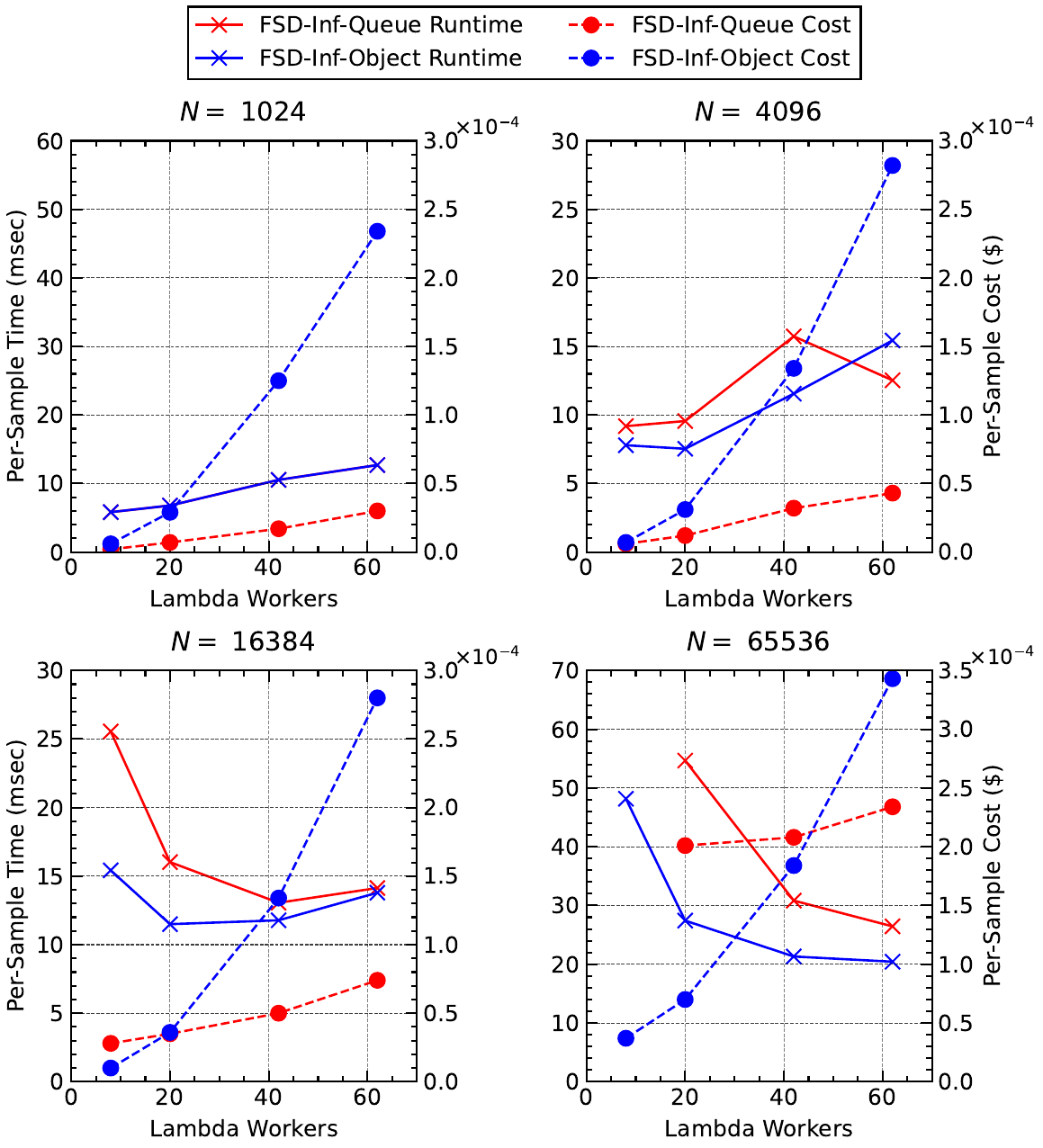}
\caption{Per-sample runtime and cost of FSD-Inf-Queue and FSD-Inf-Object.}
\label{fig:runtime-and-cost-2x2}
\end{figure}

Having established our solution as a scalable inference platform, we now explore the differences in performance between its variants (FSD-Inf-Serial, FSD-Inf-Queue, FSD-Inf-Object), as well as Sage-SL-Inf. Figure~\ref{fig:runtime-and-cost-2x2} and Table~\ref{tab:sds-only-per-sample-runtimes} show that for \textbf{smaller neuron counts} ($N = 1024, 4096$), using fewer workers is optimal both in terms of cost and performance (note that FSD-Inf-Parallel refers to the best configuration between FSD-Inf-Queue/FSD-Inf-Object). In particular, the FSD-Inf-Serial configuration, with no communication overhead, would be preferred, in line with our design recommendations in Section \ref{ss:design-recommendations}. The parallel configurations (for both FSD-Inf-Queue and FSD-Inf-Object) have similar performance profiles, although FSD-Inf-Object costs grow more quickly with increasing worker parallelism than those of FSD-Inf-Queue. As per Section \ref{s:cost-model}, this behaviour is expected when parallelism increases while data volumes are low.

We also observe that Sage-SL-Inf is outperformed by FSD-Inf-Serial for all model sizes. It should also be noted that Sage-SL-Inf was only able to perform inference with 8,000, 2,500 and 1,000 samples for $N = 1024, 4096, 16384$, respectively.

\begin{table}[h]
\centering
\caption{End-to-end per-sample runtime (ms) of optimal parallel FSD-Inference variant, FSD-Inf-Serial and Sage-SL-Inf}
\begin{tabular}{rrrr}
\hline
N & FSD-Inf-Parallel & FSD-Inf-Serial & Sage-SL-Inf \\ \hline
1024 & 6.43 & 2.00 & 2.26* \\
4096 & 8.22 & 7.88 & 10.06*\\
16384 & 12.97 & 32.62 & 37.07*\\
65536 & 23.53 & - & -\\
\hline
\end{tabular}
\label{tab:sds-only-per-sample-runtimes}
\end{table}

For $N = 16384$, increasing parallelism yields improvements in performance up to $\sim$ 20 workers (FSD-Inf-Object) and $\sim$ 40 workers (FSD-Inf-Queue), but beyond that, a slow degradation is seen. For these optimal parallelism levels, costs are similar between the FSD-Inference variants. For this $N$, parallel configurations significantly outperform FSD-Inf-Serial.

For $N = 65536$, performance consistently improves as parallelism approaches $P = 60$. At this level of parallelism, FSD-Inf-Queue costs are far lower than those of FSD-Inf-Object. Experiments with $P > 60$ then show a slow decline in performance with increasing cost (not shown). 
Neither Sage-SL-Inf nor FSD-Inf-Serial were able to perform inference with this model, due to exceeding FaaS memory limits. 
FSD-Inf-Queue was also unable to run within the maximum FaaS runtime for $P = 8$, as the extreme communication volume saturated the small number of pub-sub/queueing resources.

\subsubsection{Discussion}
We observe that FSD-Inf-Object costs increase linearly with worker parallelism $P$, and are \textit{largely independent of $N$}. As shown in Section~\ref{sss:object-cost-model}, current object storage pricing models charge only based on the number of requests, and not on the \textbf{volume} of data transferred. On the contrary, FSD-Inf-Queue costs grow much more slowly with $P$ for a given $N$, due to $\approx 1$ OOM cheaper API requests. For smaller $N$ (1024, 4096, 16384), the compressed $\bar{x}_{mn}^{k-1}$ is handled by a small number of publishes. However, for $N > 16384$, multiple publishes are increasingly required for each target at each layer, explaining the higher costs for these workloads (although, costs grow slowly with $P$ even at $N = 65536$).
These findings reinforce our recommendations in Section \ref{ss:design-recommendations}; FSD-Inf-Serial is the most performant and cost-effective solution for small models ($N = 1024, 4096$), while FSD-Inf-Queue offers an competitive performance for moderately-sized models ($N = 16384$), as well as a favourable cost profile for increasing parallelism. Finally, FSD-Inf-Object is the leading approach for large models ($N = 65536$).
As further work, these findings (with our cost model) could enable automatic runtime selection of the optimal configuration for specific workloads, given latency and cost priorities.

\subsection{Hypergraph Partitioning Performance}
\label{ss:hgp-performance}
In Table ~\ref{tab:hgp-dnn-vs-rp}, we compare the performance of our hypergraph partitioning scheme (HGP-DNN) against RP, the PaToH \cite{PaToH} random partitioning scheme, for $N = 16384$, $P = 42$. Using HGP-DNN we achieve a reduction in overall data volume sent between parallel FaaS instances of almost 1 OOM, as well as a significant improvement in per-sample runtime.


\begin{table}[h]
\centering
\caption{FSD-Inf-Object communication volumes under HGP-DNN and Random Partitioning (RP). Evaluated with $N = 16384$, $P = 42$}
\begin{tabular}{m{1.6cm} m{1.6cm} m{1.6cm} m{1.6cm}}
\hline
Partitioning Scheme & Data Volume Sent (Bytes) & NNZ Sent Per Target & Per-Sample Runtime (ms) \\ \hline
\textbf{HGP-DNN} & \textbf{3,895,079,200} & \textbf{17,888} & \textbf{11.78} \\
RP & 36,374,240,000 & 86,020 & 27.90 \\
\hline
\end{tabular}
\label{tab:hgp-dnn-vs-rp}
\end{table}

\subsection{Cost Model Validation}

To validate our cost model, we programmatically capture fine-grained metrics (51 per-layer and 26 per-batch), and use them to calculate our predicted costs. We capture actual charges via detailed AWS Cost and Usage reports.
We filter the reports to include only relevant items (i.e., communication/Lambda invocation and runtime/data transfer) in the appropriate time window, sum the charges, and compare against predicted costs.
As an example, we consider $N = 16384$, $P = 20$ (10000 samples).
\textbf{FSD-Inf-Queue}: Pred. (Comp. \$0.10, Comms. \$0.25, Total \$0.35), Actual (Comp. \$0.10, Comms. \$0.25, Total \$0.35).
\textbf{FSD-Inf-Object}: Pred. (Comp. \$0.09, Comms. \$0.28, Total \$0.37), Actual (Comp. \$0.09, Comms. \$0.28, Total \$0.37).
Note that we ran all experiments in a dedicated AWS account, to ensure no extraneous activity.
We confirmed that the real AWS costs incurred for each configuration matched those predicted by the model.

\section{Conclusion}
\label{s:conclusion}

In this work, we present FSD-Inference, the first fully serverless and highly scalable system for distributed ML inference. 
We demonstrate the suitability of fully serverless compute and communication offerings for distributed ML. 
We offer recommendations for designing high throughput/low latency serverless inference solutions, supported by a formalized and validated cost model.
We introduce optimizations to address scalability and performance challenges associated with distributed inference.
We perform a rigorous experimental evaluation of FSD-Inference and compare it with several potential alternative cloud and server-based approaches, analogous to popular ML inference solutions. 
We show that FSD-Inference offers impressive cost-effectiveness and scalability, achieving competitive performance even against optimized HPC solutions. Further, we show that publish-subscribe and queueing offerings can be adapted to offer an attractive cost profile for increasing compute parallelism, and hence are a viable alternative to object storage for FaaS inter-process communication.
Unlike provisioned systems, FSD-Inference can dynamically scale up/down to accommodate varying model sizes and sporadic inference settings. 
We further enhance the performance of FSD-Inference using hypergraph partitioning.
This further reduces communication by almost 1 order-of-magnitude when compared to alternatives, and achieves computational load balance across parallel FaaS workers. The combined model and data-parallelism enables FSD-Inference to accommodate much larger ML workloads than would otherwise be possible in the resource-constrained FaaS setting.

\section*{Acknowledgments}

This research is supported in part by the Feuer International Scholarship in Artificial Intelligence.



%




\bibliographystyle{IEEEtran}
\bibliography{sample-base}

\end{document}